\documentclass[letterpaper]{article} % DO NOT CHANGE THIS
\usepackage{aaai25}  % DO NOT CHANGE THIS
\usepackage{times}  % DO NOT CHANGE THIS
\usepackage{helvet}  % DO NOT CHANGE THIS
\usepackage{courier}  % DO NOT CHANGE THIS
\usepackage[hyphens]{url}  % DO NOT CHANGE THIS
\usepackage{graphicx} % DO NOT CHANGE THIS
\usepackage{amssymb}
\usepackage{natbib}  % DO NOT CHANGE THIS AND DO NOT ADD ANY OPTIONS TO IT
\usepackage{amsmath} 
\usepackage{caption} % DO NOT CHANGE THIS AND DO NOT ADD ANY OPTIONS TO IT
\usepackage{algorithm}
\usepackage{algorithmic}
\newcommand{\be}{\begin{equation}}
\newcommand{\ee}{\end{equation}}
\usepackage{newfloat}
\usepackage{listings}
\DeclareCaptionStyle{ruled}{labelfont=normalfont,labelsep=colon,strut=off} % DO NOT CHANGE THIS
\floatstyle{ruled}
\newfloat{listing}{tb}{lst}{}
\floatname{listing}{Listing}
\title{Intelligent IoT Attack Detection Design via ODLLM with Feature Ranking-based Knowledge Base }
\author{
Satvik Verma\textsuperscript{1}, Qun Wang\textsuperscript{1}, E. Wes Bethel\textsuperscript{1,2}
}
\affiliations{
    \textsuperscript{1}Department of Computer Science, San Francisco State University, San Francisco, CA, 94132\\

	 \textsuperscript{2}Lawrence Berkeley National Laboratory, Berkeley CA, 94720\\
    
     sverma4@sfsu.edu, claudqunwang@ieee.org, ewbethel@sfsu.edu
}

\usepackage{bibentry}
\begin{document}

\maketitle

\begin{abstract}
The widespread adoption of Internet of Things (IoT) devices has introduced significant cybersecurity challenges, particularly with the increasing frequency and sophistication of Distributed Denial of Service (DDoS) attacks. Traditional machine learning  (ML) techniques often fall short in detecting such attacks due to the complexity of blended and evolving patterns. To address this, we propose a novel framework leveraging On-Device Large Language Models (ODLLMs) augmented with fine-tuning and knowledge base (KB) integration for intelligent IoT network attack detection. By implementing feature ranking techniques and constructing both long and short KBs tailored to model capacities, the proposed framework ensures efficient and accurate detection of DDoS attacks while overcoming computational and privacy limitations. Simulation results demonstrate that the optimized framework achieves superior accuracy across diverse attack types, especially when using compact models in edge computing environments. This work provides a scalable and secure solution for real-time IoT security, advancing the applicability of edge intelligence in cybersecurity.
\end{abstract}

\section{Introduction}
The proliferation of IoT sensors in both residential and industrial environments has led to the generation of vast amounts of data that require timely and effective processing to facilitate rapid decision-making \cite{iotddos2}. However, IoT sensors are particularly vulnerable to various cyber-attacks, especially DoS and Distributed DDoS attacks. These attacks can cause significant losses and further harm, making the quick and accurate identification of such threats critically important \cite{iotddos1}.

ML algorithms have been extensively used to detect abnormal DDoS traffic. The authors in \cite{trddos1} proposed a methodology to convert the network traffic data into image form and trained a CNN model for DDoS detection. \cite{trddos2} presented a DDoS attack detection algorithm based on traffic variations and LSTM and CNN models. The authors in \cite{trddos3} utilized LSVM, Neural Network, and Decision tree to detect abnormal activities such as DDOS features. Traditional ML algorithms rely on large datasets for training and face numerous limitations in this context. Moreover, when multiple attack types are mixed together, these algorithms often struggle to perform effectively. 

The emergence of large models has shown promise in the real-time and accurate identification of various abnormal network traffic patterns \cite{llmabnormal}. Nonetheless, these models typically require substantial computational resources for training and deployment, consuming significant amounts of computing power and electricity.
Moreover, utilizing third-party models introduces data privacy and security concerns, which are particularly pertinent in network attack defense scenarios \cite{llmpry}. In distributed, large-scale IoT networks, such models often fail to meet user needs promptly due to their resource-intensive nature and potential latency issues. This has led to increased interest in edge computing paradigms, where edge intelligence and on-device large models have garnered significant attention from researchers \cite{llmqw1} \cite{oct1}.
By employing techniques like model pruning and compression, smaller models can deliver functionalities comparable to their larger counterparts in most situations \cite{llmqw1} \cite{oct2}. Building upon this, developing applications on ODLLMs can ensure affordable intelligent decision-making and local data processing \cite{qunodllm}. However, when applying ODLLMs to network attack detection, a lack of necessary background knowledge within the models necessitates fine-tuning and the integration of knowledge base (KB) to enhance their performance.

Therefore, we propose a novel system that leverages ODLLMs augmented with KB assistance to improve the detection accuracy of DDoS attacks in IoT environments. Our system addresses the challenges of computational resource constraints and privacy concerns by enabling on-device processing. We demonstrate that with appropriate KB support, ODLLMs can achieve performance comparable to larger models while operating within the limitations of edge devices.
%This approach offers a viable solution for real-time, accurate detection of DoS and DDoS attacks in distributed IoT networks, enhancing security and mitigating potential losses caused by such attacks. 
The contributions of this paper are as follows:

A novel DDoS attack detection system that utilizes ODLLMs for IoT environments is proposed.
% Address the lack of background knowledge in ODLLMs by integrating specialized KB to improve detection accuracy.
% We conduct extensive experiments to evaluate the effectiveness of our proposed system, demonstrating its capability to accurately identify complex and diverse network attacks while operating efficiently on edge devices.
% By addressing the critical need for efficient and secure network attack detection in IoT environments, our work contributes to the advancement of edge intelligence and paves the way for more robust IoT security solutions.
 We first introduce a novel approach for feature prioritization using a Random Forest Regressor (RFR) to rank the most critical features for different DDoS attack types, enabling the construction of efficient and scalable KBs. We then address the limitations of smaller ODLLMs by designing a simplified KB that retains only high-impact features, significantly improving predictive performance while reducing computational overhead. Third, we demonstrate the effectiveness of our framework through extensive experiments on the CICIoT 2023 dataset, achieving high accuracy in detecting various DDoS attack types. Our results highlight the critical role of tailored KB designs in enabling efficient and accurate attack detection on resource-constrained edge devices, paving the way for scalable and secure IoT network solutions.

The subsequent sections are organized in the following manner. The system model and problem formulation are presented in Section II. The proposed feature ranking-based KB design is developed in Section III. The findings of the simulation are presented in Section IV. Finally, Section V provides the concluding remarks for this paper.

\begin{figure}
\setlength{\abovedisplayskip}{3pt}
	\setlength{\belowdisplayskip}{3pt}
\centering
\includegraphics[width=0.90\linewidth]{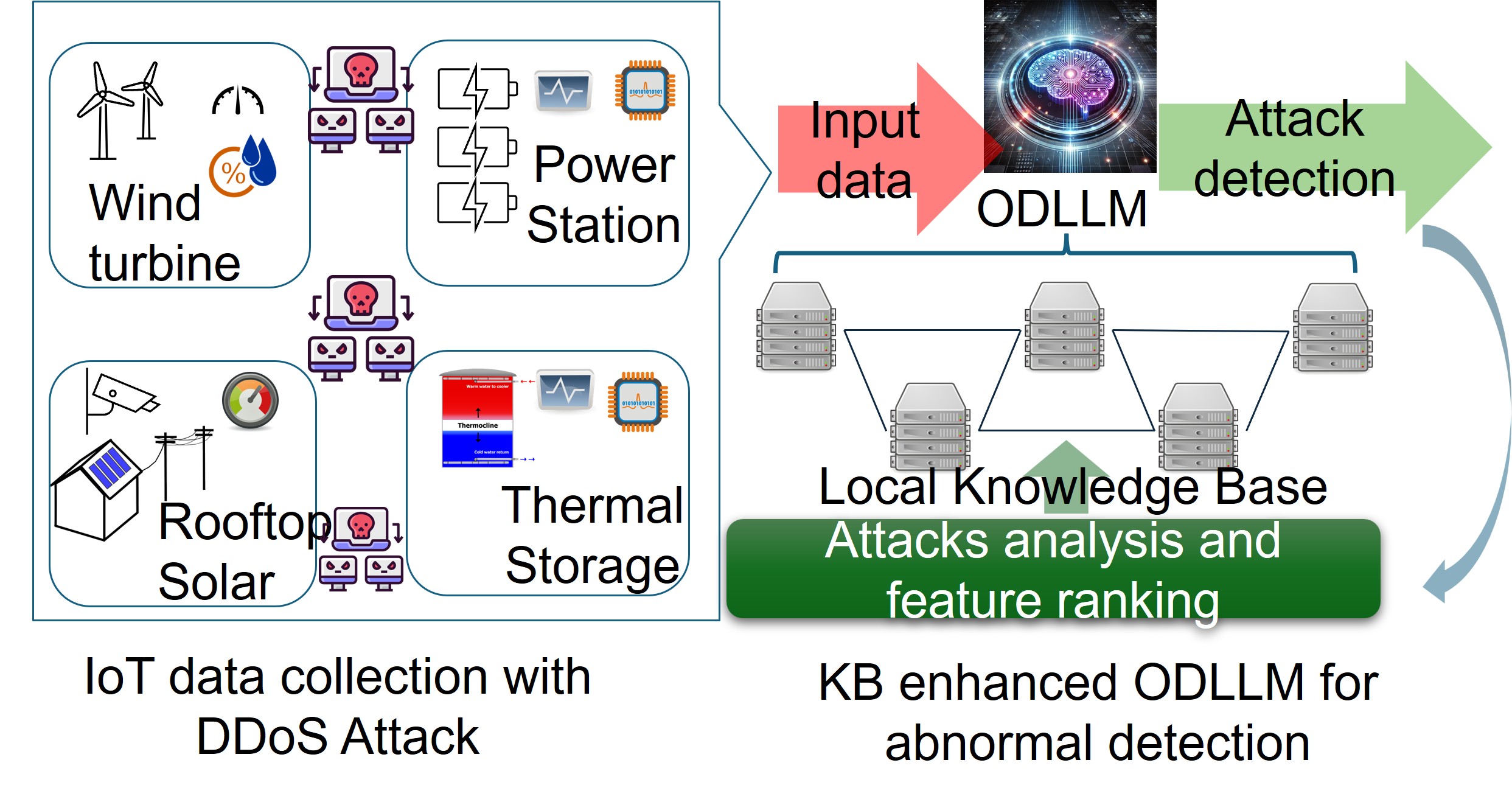}
\caption{System model.\label{sys}}
\vspace{-0.5cm}
\end{figure}	
	
\section{System Model and Problem Formulation}
%We propose an intelligent network attack detection framework based on ODLLM to enhance the security of IoT networks. The system is designed to efficiently identify and classify various types of Distributed Denial of Service (DDoS) attacks by analyzing network traffic data in real time.
\subsection{System Model}
As shown in Fig. \ref{sys}, the framework operates in three primary stages:

(1) Feature Analysis and Ranking: The system begins by extracting key features from the historical records of DDoS traffic data. These features include important values such as protocol types, packet rates, inter-arrival times, TCP flags, and statistical metrics like average packet size and variance. By analyzing these features, the system ranks them based on their significance in indicating abnormal network behavior. This prioritization allows the model to focus on the most impactful indicators of potential attacks.

(2) Knowledge Base Construction: The KB serves as the intermediary layer between feature analysis and anomaly detection. It provides a structured repository of critical insights derived from the ranked features. The KB is constructed in two formats to optimize compatibility with different ODLLM model capacities. Long KBs are used for detailed analysis with medium-size ODLLM, and short KBs are used for lightweight ODLLM applications.
The KB encapsulates the most distinguishing features of each attack type. These features are encoded as concise descriptors that facilitate quick comparison with incoming traffic data. 
%For instance, DDoS-ICMP\_Flood is characterized by large average packet sizes and high packet counts, while DDoS-TCP\_Flood exhibits high SYN counts and irregular flow durations.

% The construction process incorporates domain knowledge and statistical insights, ensuring robustness against diverse attack scenarios. Data-driven analysis of labeled datasets aids in refining the KB to include features with high discriminative power. Additionally, compact representations are prioritized to enable real-time deployment on resource-constrained edge devices.

% By aligning the KB structure with the output of the feature analysis phase, the system ensures that critical information is seamlessly transferred to the LLM for effective anomaly detection. This integration enhances the interpretability of the LLM’s decisions, providing a transparent and modular framework for IoT attack detection.

(3) Integration with LLM for Anomaly Detection: After identifying the important features and constructing KB, the system integrates them with an ODLLM to perform anomaly detection. The LLM leverages its advanced reasoning capabilities and contextual understanding to interpret the feature set comprehensively. By incorporating domain-specific knowledge, the LLM can accurately predict the type of network attack occurring. 
%This approach enables the detection of complex and subtle attack patterns that traditional machine-learning models might overlook.
%The architecture of the framework ensures that the detection process is both rapid and accurate, catering to the demands of large-scale and distributed IoT environments. 

By deploying the model on edge devices, we aim to maximize the accuracy of abnormal traffic detection.

\subsection{Types of DDoS Attack}

% DDoS attacks can cause substantial disruptions, leading to service outages, data loss, and significant financial and reputational damage.
We consider four types of DDoS attacks and their characteristics:

\textbf{ICMP Flood Attack} overwhelms the target with a high volume of ICMP echo requests, causing the network to become congested and unresponsive.
This attack usually causes increased bandwidth consumption, degraded service performance, and potential downtime.

\textbf{UDP Flood Attack} sends a large number of UDP packets to random ports on the target server, forcing it to process unnecessary requests.
It usually has random destination ports and stateless protocol exploitation.
UDP flood will increase CPU usage, and denial of legitimate service requests.

\textbf{TCP SYN Flood Attack} exploits the TCP handshake mechanism by sending numerous SYN packets without completing the handshake, consuming server resources. It usually has elevated SYN flags, half-open TCP connections, and spoofed IP addresses. It will exhaust connection tables, leading to an inability to establish new legitimate connections.

\textbf{TCP PSH+ACK Flood Attack} sends a large number of TCP packets with the PSH (Push) and ACK (Acknowledgment) flags set to overwhelm the target's processing capabilities. It usually involves high volumes of PSH and ACK packets that mimic normal traffic patterns, making them difficult to filter. This flood increases processing overhead, leads to resource depletion, and can cause potential service crashes.

% \textbf{TCP RST/FIN Flood Attack:} Sends numerous TCP packets with the RST (Reset) or FIN (Finish) flags to disrupt active connections and cause session terminations. It typically features elevated RST and FIN flags that mimic normal session endings, resulting in abrupt connection closures. This attack causes connection resets, loss of data transmission integrity, and service instability.

% \textbf{DDoS-Synonymous IP Flood:} Involves multiple source IP addresses sending packets to the target simultaneously, complicating detection and mitigation efforts. It usually features diverse IP sources and high-volume traffic, often utilizing botnets. This flood increases difficulty in filtering malicious traffic, overwhelms network infrastructure, and leads to extensive resource consumption.

\subsection{Problem Formulation}

Our objective is to design a DDoS attack detection model that maximizes the accuracy of identifying various attacks in IoT environments while operating efficiently on edge devices with limited computational resources.
Let $\mathcal{D} = \{ (\mathbf{x}_i, y_i) \}_{i=1}^N$ be the traffic dataset, where
 $\mathbf{x}_i \in \mathbb{R}^d$ represents the feature vector extracted from network traffic data for the $i$-th sample.
 $y_i \in \mathcal{Y} = \{1, 2, \dots, C\}$ is the corresponding label indicating the attack type, with $C$ being the number of attack classes (including normal traffic).
Let $f_{\theta} : \mathbb{R}^d \rightarrow \mathcal{Y}$ be the detection ODLLM model enhanced by KB $\theta$, which maps input features to predicted labels.

The primary goal is to find the optimal KB $\theta^*$ that maximizes the overall accuracy on the dataset $\mathcal{D}$.
The accuracy $\mathcal{A}(\theta)$ can be defined as:
\be
\mathcal{A}(\theta) = \frac{1}{N} \sum_{i=1}^N \delta\left( f_{\theta}(\mathbf{x}_i), y_i \right),
\vspace{-0.1cm}
\ee
where $\delta(a, b)$ is the Kronecker delta function:
\be
\delta(a, b) = \begin{cases}
1, & \text{if } a = b, \\
0, & \text{if } a \neq b.
\end{cases}
\vspace{-0.1cm}
\ee
The problem can be formulated as an optimization problem:
\be
\theta^* = \arg\max_{\theta} \mathcal{A}(\theta).
\vspace{-0.1cm}
\ee

Considering the edge computing resource limitations and ODLLM model constraints, we need to design KB to maximize the detection accuracy.

\section{Feature Ranking and Knowledge Base Design}

To build a reliable model capable of differentiating between various types of attacks, we implemented a structured feature ranking and KB development methodology. This process involved selecting high-impact features, constructing an adaptable KB, iterating for accuracy improvements, and tailoring our approach based on model capacity.

\subsection{Feature Ranking with Random Forest Regressor}

% Our initial step in building the knowledge base was to rank features by their significance for each attack type. We applied a Random Forest Regressor (RFR) model, a robust feature-ranking method that helps identify features with the highest predictive value for classification. The RFR model analyzed the dataset, which contained various attack types, and provided an ordered list of features for each attack. We selected the top ten features for each attack type, focusing on characteristics that could act as clear differentiators between attack patterns.

% For each of the top features, we extracted descriptive statistics, including the upper and lower bounds and the median value. This range of values allowed us to understand typical and boundary behavior for each feature, creating a numerical profile that captured the unique signature of each attack. For example, features such as protocol type, packet size, and inter-arrival time (IAT) often varied across attack types and provided critical insights for differentiating behaviors. These bounded values became the foundation for structuring a clear and precise knowledge base that could inform the model’s predictions.

%===========================================================
Our initial step in constructing the KB involved ranking features by their importance for each attack type. We employed a Random Forest Regressor (RFR) to evaluate feature significance.
%a robust ensemble learning method that evaluates feature significance based on their contribution to reducing the mean squared error (MSE) during training. 
The RFR assigns importance based on how well each feature helps split the data to classify attacks correctly. The more a feature contributes to reducing classification mean squared error (MSE) across multiple decision trees, the higher its importance \cite{rfr}.
The importance for feature $\alpha_i$ is computed as:
\be
I(\alpha_i) = \frac{1}{T} \sum_{t=1}^{T} \Delta \text{MSE}_t(\alpha_i),
\vspace{-0.1cm}
\ee
where $T$ is the total number of decision trees in the forest, and $\Delta \text{MSE}_t(\alpha_i)$ represents the reduction in MSE for tree $t$ when splitting on feature $\alpha_i$.
The RFR model analyzed a labeled dataset consisting of various attack types, producing an ordered list of features ranked by their importance score $I(\alpha_i)$. From this ranking, we selected the top $k = 10$ features for each attack type, focusing on characteristics with the highest predictive value for distinguishing between attack patterns.
For each selected feature, we extracted descriptive statistics, including the {lower bound} $\min(\alpha)$, {upper bound} $\max(\alpha)$, and the {median value} $\text{med}(\alpha)$. These statistics defined a feature range:
\be
\text{Range}(\alpha_i) = [\min(\alpha_i), \max(\alpha_i)],
\ee
which encapsulates typical and boundary behavior for feature $\alpha_i$. The descriptive statistics for each $\alpha_i$ are given as:
\be
\text{Statistics}(\alpha_i) = \{\min(\alpha_i), \text{med}(\alpha_i), \max(\alpha_i)\}.
\ee
This range and statistical profile provided a numerical signature for each attack type, effectively differentiating between behaviors such as protocol type, packet size, and inter-arrival time (IAT). 
%For example, {DDoS-ICMP\_Flood} attacks typically exhibited large packet sizes ($\text{Range}(\text{Packet Size}) = [42, 30,329]$), while {DDoS-UDP\_Flood} attacks had high inter-arrival times ($\text{Range}(\text{IAT}) = [0, 167,639,426]$).
These bounded values formed the foundation for a structured and precise KB, enabling the model to learn the unique signatures of each attack type and improve prediction accuracy.

We consider the dataset CICIOT 2023, which is a comprehensive benchmark dataset designed for evaluating intrusion detection systems in IoT environments, featuring diverse network traffic types, including normal and malicious behaviors, across multiple IoT protocols and attack scenarios \cite{ciciot2023}.
\subsubsection{DDoS ICMP Flood}

\begin{figure}
\setlength{\abovedisplayskip}{3pt}
	\setlength{\belowdisplayskip}{3pt}
\centering
\includegraphics[width=0.47\textwidth]{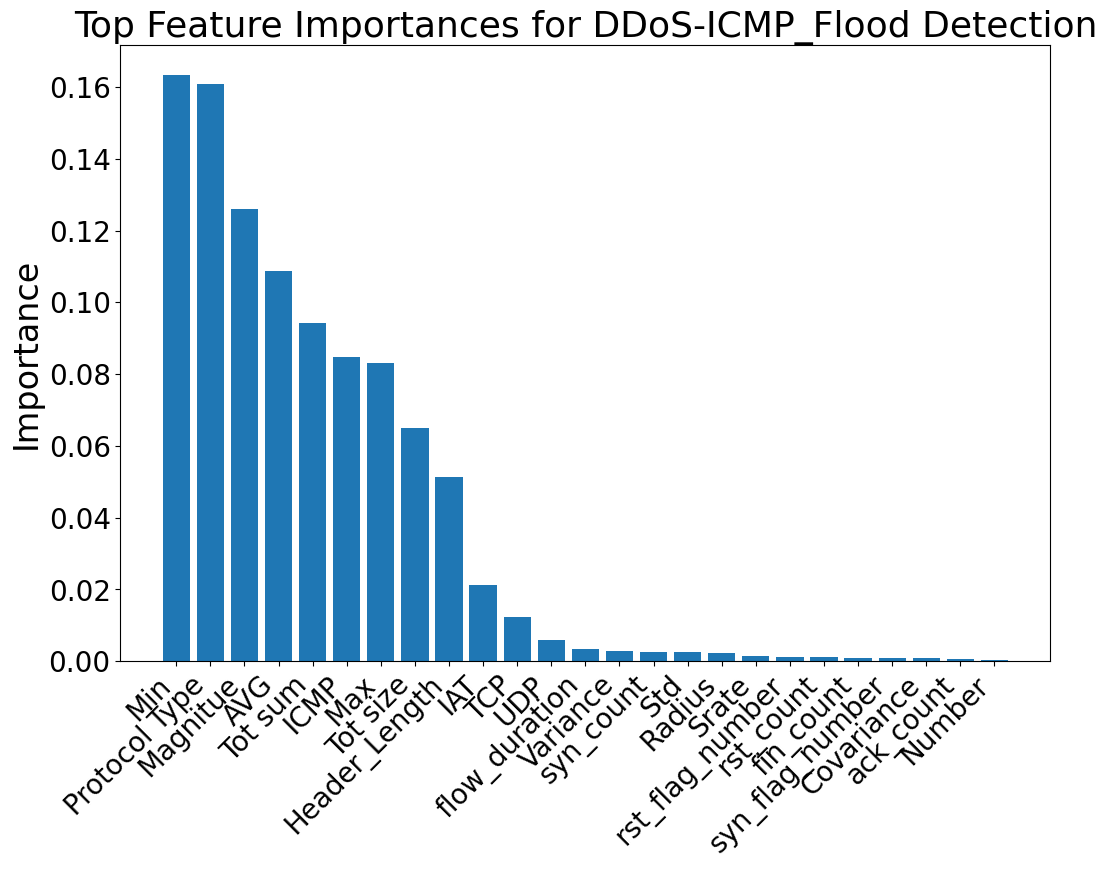}
\caption{Ranked features for DDoS ICMP Flood attack using Random Forest Regressor on the CIC IoT 2023 Dataset.}
\label{fig:icmp_features}
\vspace{-0.5cm}
\end{figure}

The feature ranking for  DDoS-ICMP\_Flood attack is shown in Fig \ref{fig:icmp_features}. The top two features identified for DDoS-ICMP\_Flood attacks in our dataset are the \textit{MIN} value and \textit{Protocol Type}. The \textit{MIN} value represents the minimum size of network traffic packets and exhibits distinct patterns during ICMP flood attacks. This is because ICMP flood attacks often involve numerous small packets, with the \textit{MIN} value ranging from 42.0 to 992.72 and a median of 42.0. In over 99\% of cases, the \textit{MIN} value is exactly 42.0, which is consistency as a signature feature for ICMP floods.
The \textit{Protocol Type} is another high-importance feature, as ICMP floods specifically utilize the Internet Control Message Protocol (ICMP). The \textit{Protocol Type} ranges from 0.77 to 15.35, with a median of 1.0, which aligns with the expected value for ICMP packets according to standard protocol definitions. 
%This feature plays a pivotal role in differentiating ICMP flood attacks from other types of DDoS attacks that rely on different protocols, such as UDP or TCP.

Additionally, features like \textit{Magnitude} and \textit{ICMP} were observed to have moderate importance. The \textit{Magnitude} feature, which quantifies the volume or intensity of network traffic, is particularly indicative of DDoS attacks as ICMP floods typically generate high bursts of traffic. In our dataset, the \textit{Magnitude} ranges from 9.16 to 59.79, with a median of 9.16, demonstrating that most ICMP flood attacks are characterized by relatively high traffic intensity. Similarly, the \textit{ICMP} feature, reflecting the presence of ICMP packets in the traffic, serves as a direct identifier for this attack type. Its values range from 0.0 to 1.0, with a median of 1.0, indicating that ICMP packets are consistently present in traffic associated with ICMP flood attacks.
Protocol-specific features, such as HTTP, DNS, SSH, and flag numbers, were found to have little to no relevance for identifying ICMP flood attacks. These features consistently showed zero or near-zero importance scores, as they are unrelated to the characteristics of ICMP floods.

% Therefore, features of high importance directly contribute to distinguishing between attack and non-attack traffic. These include: Protocol Type, ICMP Presence,
% Protocol Type: Essential for identifying ICMP-specific traffic.
% ICMP Presence: Indicates the direct involvement of ICMP packets.
% Total Size (Tot size): Highlights the aggregate packet size during the attack.
% Medium-Importance Features: Capture patterns indicative of DDoS attacks but are not exclusive to ICMP floods. Examples include:
% MIN and MAX Values: Representing packet size variations.
% Total Sum (Tot sum): Reflecting the cumulative impact of traffic flow.
% Low or Zero-Importance Features: Unrelated or redundant in identifying ICMP floods. For instance:
% Protocol-Specific Metrics (e.g., SMTP, HTTP): Irrelevant as these protocols are not involved in ICMP floods.
% Flag Numbers: Provide no additional discriminatory information for ICMP-specific attacks.
This feature ranking allows the model to utilize KB to focus on the most relevant characteristics for accurate detection of DDoS-ICMP\_Flood attacks. The KB can be given as:

\begin{lstlisting}
If the attack is DDoS ICMP flood, it should exhibit the following characteristics:
- Protocol Type: Has to be 1.0 for ICMP.
- ICMP Indicator: Has to be 1.0 for ICMP.
- Min Packet Size: Ranges from 42.0 to 992.72, commonly at 42.0.
- Magnitude: Intensity ranges from 9.17 to 59.80, with a typical value near 9.17.
- Average Packet Size (AVG): Spans from 42.0 to 1885.5, often around 42.0.
- Total Sum of Packets (Tot sum): Between 42.0 and 19764.8, commonly near 441.0.
- Max Packet Size: Has to be around 42.0.
- Total Size of Packets (Tot size): Has to be 42.0.
- Inter-Arrival Time (IAT): Very high, between 0.0 and 100179851.34, with a median around 83128994.35.
\end{lstlisting}

\subsubsection{DDoS UDP Flood}

\begin{figure}
\setlength{\abovedisplayskip}{3pt}
	\setlength{\belowdisplayskip}{3pt}
\centering
\includegraphics[width=0.47\textwidth]{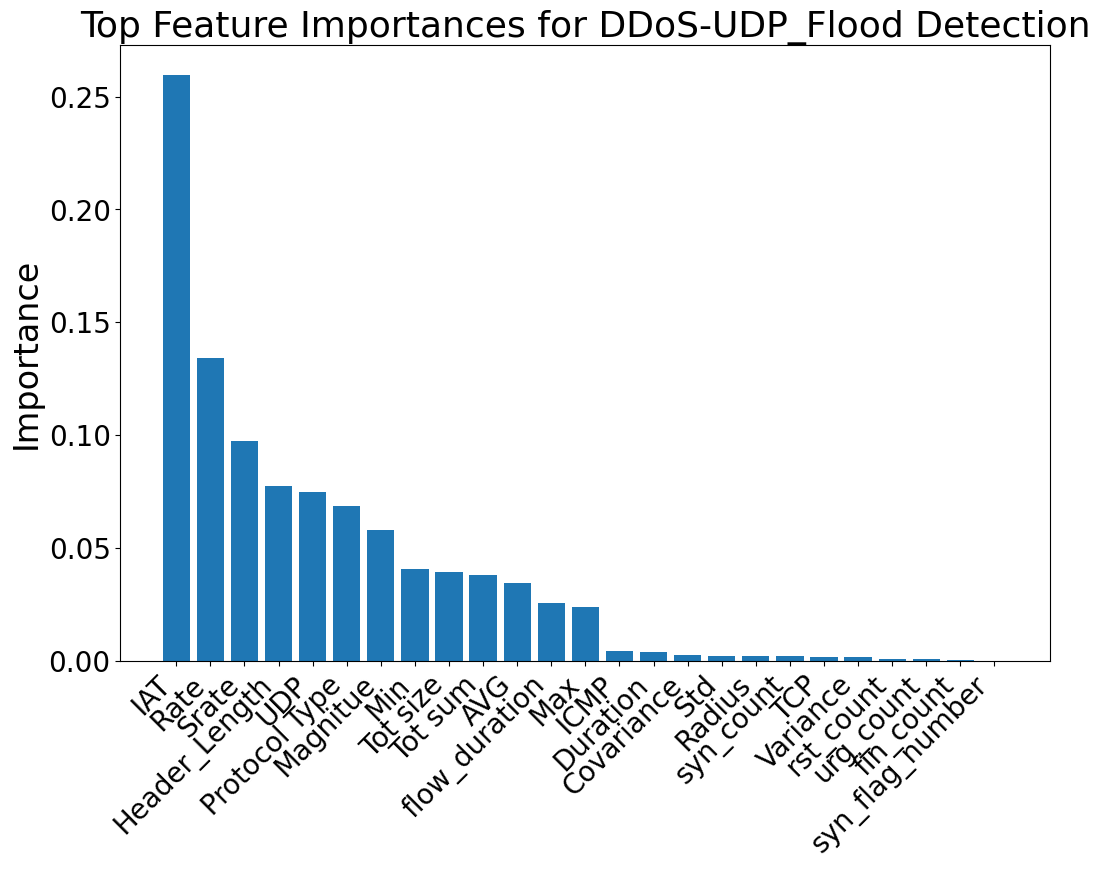}
\caption{Ranked features for DDoS UDP Flood attack using Random Forest Regressor on the CIC IoT 2023 Dataset.}
\label{fig:udp_features}
\vspace{-0.5cm}
\end{figure}

The feature ranking of DDoS-UDP\_Flood attacks is shown in Fig. \ref{fig:udp_features}. 
The two most critical features are \textit{IAT} and \textit{Rate}. This is because of the nature of UDP floods, which are characterized by bursts of packets with minimal inter-arrival time and a high packet rate. The \textit{IAT} ranges from $4.3 \times 10^{-6}$ to $99,748,506.4$, with a median value of $83,102,993.46$, reflecting the rapid packet generation typical of these attacks. Similarly, the \textit{Rate} feature, which captures the volume of UDP packets sent over the network, spans from 6.0 to $1,569,352.1$, with a median of $7,480.80$. These values highlight the high-frequency, high-volume characteristics of UDP flood attacks.

Features with moderate importance include {Source Rate (Srate)}, {Header Length}, \textit{UDP}, and \textit{Protocol Type}:
\textit{Srate} reflects the source-side packet transmission rate, ranging from 6.0 to $1,569,352.1$, aligning with the traffic burst patterns typical of UDP floods.
{Header Length}, varying between 751.5 and $1,076,354.07$, represents packet sizes associated with UDP flood traffic.
The \textit{UDP} feature confirms the protocol type, typically close to 1.0, indicating the attack's reliance on the UDP protocol.
\textit{Protocol Type}, ranging from 4.84 to 17.0 with a median of 17.0, differentiates UDP floods from other attacks, as the value corresponds to the UDP protocol in networking standards.

Features with lower importance include \textit{Magnitude}, \textit{Min}, \textit{Tot Size}, and \textit{Tot Sum}.  \textit{Magnitude} represents the intensity of the traffic flow, reflecting moderate importance in identifying the attack's burst characteristics. It ranges from 9.97 to 41.16, with a median value of 10.0.
\textit{Min}, \textit{Tot Size}, and \textit{Tot Sum} play supporting roles in detecting anomalies associated with UDP floods by capturing packet-level traffic metrics,
Features such as \textit{ICMP}, \textit{TCP}, and flag numbers have minimal or zero importance for DDoS-UDP\_Flood detection.
%Additionally, protocol-specific features such as {HTTP}, {DNS}, {SSH}, and {SMTP} are unrelated to UDP-based traffic and hold no relevance for classification. This is consistent with the nature of UDP flood attacks, which do not rely on these protocols or flag characteristics, further narrowing the focus to UDP-specific behaviors.

This analysis emphasizes the role of high-importance features like {\textit{IAT}} and \textit{Rate} in distinguishing DDoS-UDP\_Flood traffic. %supported by protocol-specific insights that ensure accurate classification and reduced false positives. 
By leveraging these prioritized features, the KB can be given as:
\begin{lstlisting}
    If the attack is DDoS UDP flood, it should exhibit the following characteristics:
- Protocol Type: Close to 17.0, corresponding to the UDP protocol.
- UDP Indicator: Must be 1.0, confirming the presence of UDP packets."
- Inter-Arrival Time (IAT): Extremely varied, ranging from 4.39e-06 to 99748506.47, with a typical value around 83102993.47, reflecting high-frequency bursts.
- Rate and Source Rate (Srate): Both range from 6.01 to 1569352.19, with a common value near 7480.80, indicating high packet transmission volumes.
- Magnitude: Represents traffic intensity, ranging from 9.97 to 41.16, typically about 10.0.
- Minimum Packet Size (Min): Between 48.74 and 468.37, commonly close to 50.0, reflecting packet-level characteristics.
- Total Packet Size (Tot size): Spans from 49.88 to 1075.46, with a frequent value near 50.0.
- Total Sum of Packets (Tot sum): Ranges from 150.0 to 11576.45, with a typical value around 525.0, capturing the cumulative packet behavior.
\end{lstlisting}

\subsubsection{DDoS TCP Flood}

\begin{figure}
\setlength{\abovedisplayskip}{3pt}
	\setlength{\belowdisplayskip}{3pt}
\centering
\includegraphics[width=0.47\textwidth]{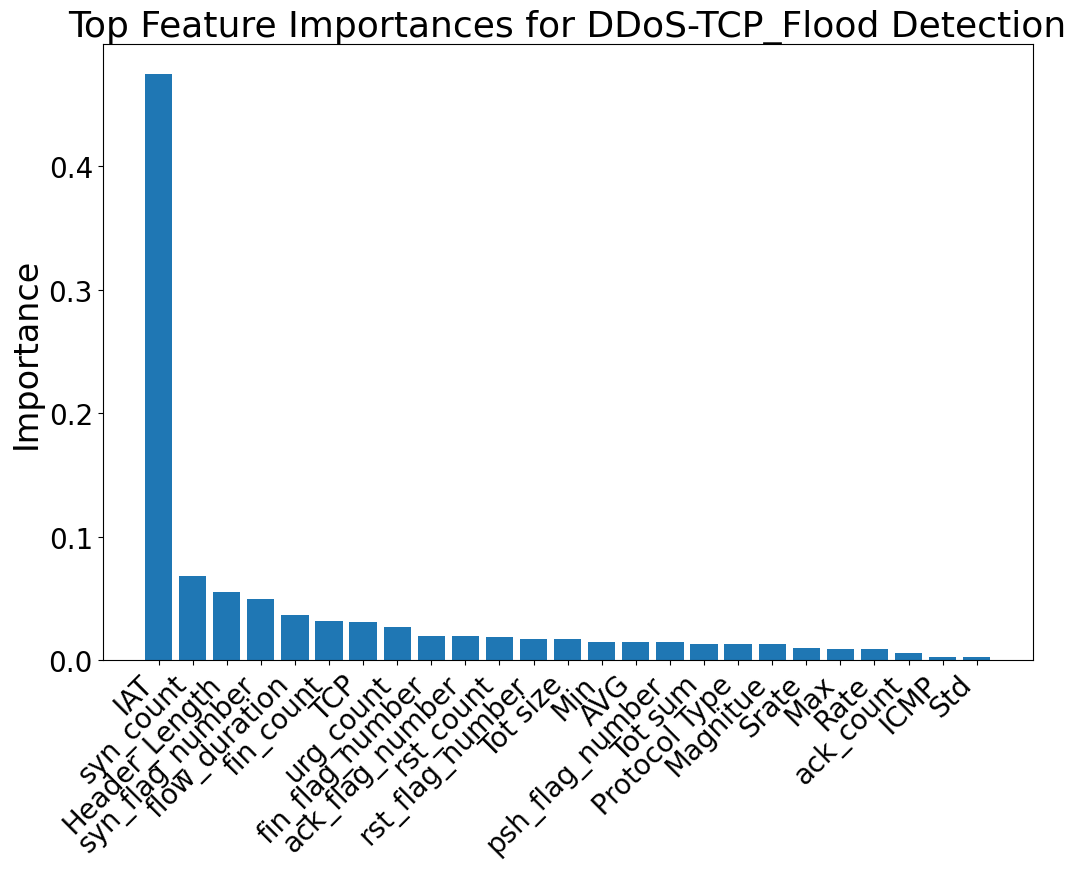}
\caption{Ranked features for DDoS TCP Flood attack using Random Forest Regressor on the CIC IoT 2023 Dataset.}
\label{fig:tcp_features}
\vspace{-0.5cm}
\end{figure}

The ranked features for DDoS-TCP\_Flood detection are illustrated in Fig. \ref{fig:tcp_features}. The two most critical features are \textit{IAT} and \textit{SYN Count}, which are essential for identifying the unique traffic patterns associated with TCP flood attacks. TCP floods often involve high-frequency traffic and repeated SYN packets aimed at exhausting server resources.
The \textit{IAT} value, which ranges from $1.3 \times 10^{-7}$ to $99,691,821.6$, with a median of $83,068,279.06$, highlights the rapid and irregular timing patterns typical of high-volume TCP traffic during an attack. Similarly, the \textit{SYN Count}, which captures the frequency of SYN packets, ranges from 0.0 to 2.25, with a median of 0.00. This low median value reflects repeated connection attempts characteristic of TCP flood behavior, where malicious actors send SYN packets to initiate multiple, incomplete TCP connections.

Features with moderate importance include \textit{Header Length}, \textit{SYN Flag Number}, \textit{Flow Duration}, and \textit{FIN Count}.
\textit{Header Length}, ranging from 50.96 to $1,264,522.69$, represents the variability in packet sizes during a TCP flood.
 \textit{SYN Flag Number}, with values typically close to 0.0, indicates a low presence of SYN flags in some TCP flood patterns.
\textit{Flow Duration}, spanning 0.0 to $1,270.91$ seconds, provides insights into the connection longevity and stability during the attack.
 \textit{FIN Count}, ranging from 0.0 to 0.45, captures the frequency of FIN packets, shedding light on TCP session termination behaviors during an attack.

%Features with lower importance, such as \textit{URG Count}, \textit{ACK Flag Number}, \textit{RST Count}, and \textit{Tot Size}, also contribute to classification. \textit{URG Count}, with values between 0.0 and 2984.6, reflects the urgency flag’s presence in some attack scenarios. \textit{ACK Flag Number} and \textit{RST Count} capture acknowledgment and reset packets, providing context for TCP traffic behavior. \textit{Tot Size}, representing the total size of packets, offers additional indicators of traffic anomalies.

% Certain features, such as \textit{ICMP}, \textit{UDP}, and application-layer protocols (e.g., \textit{HTTP}, \textit{DNS}, \textit{SMTP}), are unrelated to TCP flood attacks and consistently show minimal or zero importance. This aligns with the nature of TCP flood attacks, which exclusively involve the TCP protocol and its corresponding flag characteristics.

%This analysis underscores the significance of high-importance features like \textit{IAT} and \textit{SYN Count}, which are directly tied to the primary mechanisms of TCP floods, supported by moderate-importance features that provide additional context for accurate classification. By focusing on these features, the detection system can effectively identify and mitigate DDoS-TCP\_Flood attacks.
Based on the above feature ranking analysis, the KB can be constructed as:
\begin{lstlisting}
    If the attack is DDoS TCP flood, it should exhibit the following characteristics:
    - Protocol Type: Close to 6.0, corresponding to the TCP protocol.
    - PSH Flag Number: Should be 0.0, reflecting minimal push flags in typical TCP flood behavior.
    - TCP Indicator: Often 1.0, confirming the use of the TCP protocol.
    - URG Count: Typically 0.0, indicating no urgency flags in normal TCP traffic.
    - SYN Flag Number: Typically 0.0, showing the absence or minimal use of SYN flags in regular traffic.
    - Flow Duration: Ranges from 0.0 to 1270.90 seconds, often 0.0 in shorter-lived connections characteristic of flood traffic.
    - FIN Count: Typically 0.0, but can reach up to 0.45 in some TCP exchanges.
    - ACK Flag Number: Mostly 0.0, indicating limited acknowledgment flags in standard TCP flood traffic.
\end{lstlisting}

% This ranking highlights the importance of traffic timing, packet-level metrics, and protocol-specific flags in accurately classifying DDoS-TCP\_Flood attacks. The consistent patterns of TCP traffic, such as SYN-based flooding and abnormal session characteristics, make these features critical for effective detection.

\subsubsection{DDoS PSHACK Flood}

\begin{figure}
\setlength{\abovedisplayskip}{3pt}
	\setlength{\belowdisplayskip}{3pt}
\centering
\includegraphics[width=0.47\textwidth]{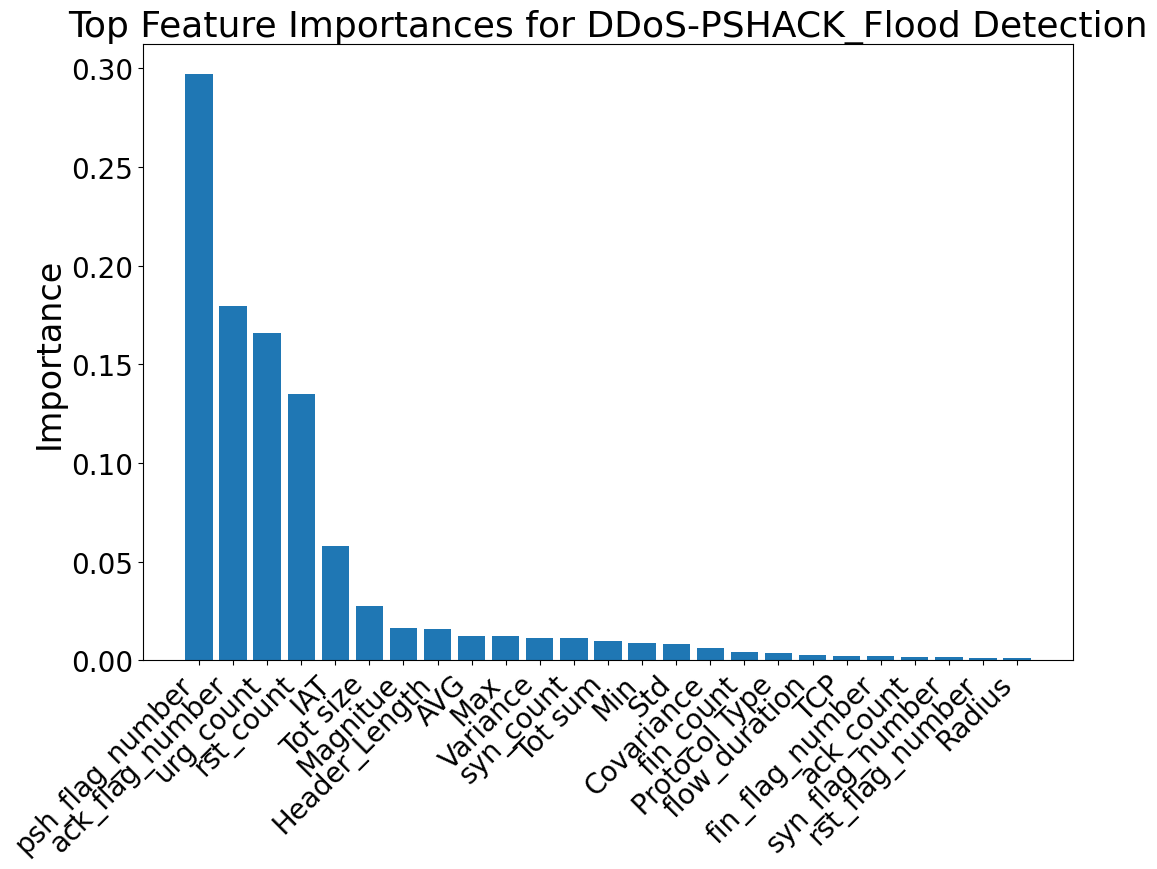}
\caption{Ranked features for DDoS PSHACK Flood attack using Random Forest Regressor on the CIC IoT 2023 Dataset.}
\label{fig:pshack_features}
\vspace{-0.5cm}
\end{figure}

The ranked features for DDoS-PSHACK\_Flood detection are illustrated in Fig. \ref{fig:pshack_features}. The two most critical features are \textit{PSH Flag Number} and \textit{ACK Flag Number}, which are integral to the attack's mechanism. PSHACK floods heavily rely on the consistent presence of PSH and ACK flags to overwhelm the target system.
The \textit{PSH Flag Number}, ranging from 0.0 to 1.0 with a median of 1.0, reflects the frequent use of PSH flags in attack packets. This consistency highlights the attack's strategy of forcing the target system to process data packets immediately. Similarly, the \textit{ACK Flag Number}, also ranging from 0.0 to 1.0 with a median of 1.0, underscores the importance of acknowledgment packets in the attack, which maintain the flood of connections and disrupt normal operations.

Other features with moderate importance include \textit{URG Count}, \textit{RST Count}, \textit{Inter-Arrival Time (IAT)}, and \textit{Total Packet Size (Tot Size)}:
 \textit{URG Count}, with values between 0.0 and 214.22 and a median of 1.0, reflects the occasional presence of urgency flags in PSHACK packets, adding to the attack's complexity.
\textit{RST Count}, ranging from 0.0 to 472.02 with a median of 1.0, indicates the frequent use of reset flags, a common tactic in PSHACK floods to disrupt TCP sessions.
\textit{IAT}, spanning $1.5 \times 10^{-5}$ to $99,998,229.54$ with a median of $83,318,215.97$, reflects the timing patterns of high-frequency bursts typical of this attack.
\textit{Tot Size}, ranging from 53.76 to 689.69 with a median of 54.0, provides additional indicators of packet size consistency in the attack.

Features with lower importance include \textit{Magnitude}, \textit{Header Length}, \textit{Average Packet Size (AVG)}, and \textit{Maximum Packet Size (Max)}:
\textit{Magnitude}, ranging from 10.34 to 31.17 with a median of 10.39, indicates the intensity of the traffic flow.
\textit{Header Length}, varying between 51.3 and $1,601,755.99$ with a median of 54.0, captures packet-level details.
\textit{AVG} and \textit{Max} values, both with medians of 54.0, emphasize the uniformity of packet sizes during the attack.

%Certain features, such as \textit{ICMP}, \textit{UDP}, and application-layer protocols (e.g., HTTP, DNS, SMTP), consistently show minimal or zero importance. These features are unrelated to PSHACK floods, which exclusively exploit TCP flags and behaviors without involving other protocols.

This analysis highlights the role of high-importance features such as PSH and ACK flags, which directly correlate with the attack's strategy, supported by moderate-importance features like RST Count and Tot Size that add context to the classification. By focusing on these key features, the detection system can effectively identify and mitigate DDoS-PSHACK\_Flood attacks. Thus, the KB is constructed as:

\begin{lstlisting}
'DDoS-PSHACK_Flood': (
    If the attack is DDoS PSHACK flood, it should exhibit the following characteristics:
    - PSH Flag Number: Must be 1.0, indicating the presence of single push flags in the traffic.
    - ACK Flag Number: Often 1.0, but can occasionally be 0.0, distinguishing it from other TCP floods.
    - URG Count: Typically 1.0 but can reach up to 367.51, reflecting the occasional use of urgency flags.
    - RST Count: Usually 1.0, highlighting the frequent use of reset flags in the attack.
    - Inter-Arrival Time (IAT): Ranges from 1.50e-05 to 99998229.53, with a common value around 83318215.96, indicating high-frequency bursts.
    - Total Packet Size (Tot size): Between 53.76 and 1177.9, typically around 54.0, showing consistent packet sizes.
    - Magnitude: Varies in intensity from 10.33 to 40.65, with a common value near 10.39.
    - Average Packet Size (AVG): Ranges from 53.34 to 1079.47, often close to 54.0, showing consistent averages.
    - Maximum Packet Size (Max): Spans from 53.76 to 3022.11, with typical values around 54.0.
)
\end{lstlisting}

\subsection{Introducing Key Features for Targeted Predictions}
To enhance accuracy further, we introduced a "key feature set" alongside the KB. This set consisted of the most discriminative features for each attack type which we got while ranking the features and then comparing the different attacks, serving as a concise reference for the model during predictions. By focusing on these critical features, the model could prioritize the most relevant characteristics before consulting the broader KB. This two-tiered structure provided both a high-level guide for classification and detailed descriptions for refining distinctions between attack types.

\subsection{Challenges with Smaller Models and KB Simplification}

% We observed significant challenges when using smaller models, such as LLaMA 3.2 3B and Phi3 Mini 3.8B. While larger models like LLaMA 3.1 8B could effectively utilize the comprehensive KB, smaller models often struggled with the volume and complexity of information. This limitation resulted in poorer predictive performance compared to using no KB at all.

% We hypothesized that smaller models, due to their reduced parameter capacity, were overwhelmed by the extensive multi-feature input in the KB. To address this, we constructed a simplified KB specifically for smaller models. This version retained only the highest-impact features for each attack type, focusing on the most distinctive characteristics while omitting secondary or redundant details. This simplification allowed smaller models to process the KB more effectively and improved their predictive accuracy. For example: 

Smaller models, such as LLaMA 3.2 3B and Phi3 Mini 3.8B, presented significant challenges in utilizing the comprehensive KB. These models struggled to process the volume and complexity of multi-feature input due to their limited parameter capacity, leading to poorer predictive performance compared to scenarios where no KB was used \cite{qunodllm} \cite{sllm}.

To address these limitations, we hypothesized that smaller models were overwhelmed by the extensive KB, which included redundant and low-impact features. As a solution, we designed a simplified KB tailored specifically for smaller models. This version focused exclusively on the highest-impact features for each attack type, retaining only the most distinctive characteristics and omitting secondary details. This streamlined approach allowed smaller models to process the KB more effectively and significantly improved predictive accuracy. For example, a simplified KB for an ICMP Flood attack can be given as:

% \begin{lstlisting}
% \textit{DDoS-ICMP\_Flood:} Protocol: ICMP; High packet rate; Low Inter-Arrival Time (IAT).
%  \textit{DDoS-UDP\_Flood:} Protocol: UDP; High packet rate; Low IAT.
%  \textit{DDoS-TCP\_Flood:} Protocol: TCP; High packet rate; Elevated SYN flag.
%  \textit{DDoS-PSHACK\_Flood:} Elevated PSH and ACK flags.
% \textit{DDoS-SYN\_Flood:} Elevated SYN flag.
%  \textit{DDoS-RSTFIN\_Flood:} Elevated RST and FIN flags.
%  \textit{DDoS-SynonymousIP\_Flood:} Multiple source IPs; High SYN counts.
% \end{lstlisting}
\begin{lstlisting}
DDoS-ICMP_Flood: Protocol: ICMP; High packet rate; Low Inter-Arrival Time (IAT).
DDoS-UDP_Flood: Protocol: UDP; High packet rate; Low IAT.
DDoS-TCP_Flood: Protocol: TCP; High packet rate; Elevated SYN flag.
DDoS-PSHACK_Flood: Elevated PSH and ACK flags.
DDoS-SYN_Flood Elevated SYN flag.
DDoS-RSTFIN_Flood: Elevated RST and FIN flags.
DDoS-SynonymousIP_Flood: Multiple source IPs; High SYN counts.
\end{lstlisting}
The simplified KB was embedded into natural language prompts to help models identify attack types more efficiently. For instance:
\begin{lstlisting}
Network Traffic Data:
- Protocol Type: TCP
- Packet Rate: 450 packets/sec (High)
- Inter-Arrival Time (IAT): Low
- TCP Flags:
    - SYN: Elevated
    - PSH: Normal
    - ACK: Normal
    - RST: Normal
    - FIN: Normal

Based on the knowledge base, determine the most likely attack type. from the following list:( DDoS-ICMP_Flood, DDoS-UDP_Flood, DDoS-TCP_Flood, DDoS-PSHACK_Flood, DDoS-SYN_Flood, DDoS-RSTFIN_Flood, DDoS-SynonymousIP_Flood, Unknow, Normal.
\end{lstlisting}
This approach successfully bridged the gap in performance for smaller models, enabling them to leverage a streamlined KB and maintain classification accuracy with minimal computational overhead.

% Our approach to knowledge base construction and refinement was guided by feature analysis, iterative improvement, and adaptation to model capacities. For larger models, the detailed KB enabled nuanced and accurate predictions. For smaller models, a simplified KB ensured focused and efficient predictions without overwhelming their capacity. This dual strategy of detailed and simplified KB versions allowed us to optimize performance across different model sizes, ensuring high accuracy in detecting and differentiating attack types.

% \section{Findings and Discussions}

% We are going to divide these Findings and Discussions into 3 sub-sections to talk compare our 3 different implementations, Base model, with large KB and with short KB, across different configurations of models, and how they compare against each other on the basis of our performance metric which is accuracy.

    \begin{table*}[h!]
\centering
\begin{tabular}{|l|c|c|c|c|c|c|c|c|c|}
\hline
\textbf{Attack Type} & \multicolumn{3}{c|}{\textbf{Llama 3.1 8B}} & \multicolumn{3}{c|}{\textbf{Phi3 Medium 14B}} & \multicolumn{3}{c|}{\textbf{Gemma2 9B}} \\ \hline
                     & \textbf{No KB} & \textbf{Long KB} & \textbf{Short KB} & \textbf{No KB} & \textbf{Long KB} & \textbf{Short KB} & \textbf{No KB} & \textbf{Long KB} & \textbf{Short KB} \\ \hline
\textbf{ICMP}   & 97.80\% & 100.00\% & 83.80\% & 50.40\% & 42.40\% & 27.40\% & 20.40\% & 100.00\% & 20.00\% \\ \hline
\textbf{UDP}    & 56.40\% & 86.60\%  & 76.60\% & 39.20\% & 31.40\% & 59.80\% & 1.80\%  & 100.00\% & 48.80\% \\ \hline
\textbf{TCP}    & 77.40\% & 3.60\%   & 77.80\% & 10.60\% & 16.80\% & 6.00\%  & 0.00\%  & 0.00\%   & 0.00\%  \\ \hline
\textbf{PSHACK} & 3.20\%  & 59.40\%  & 54.80\% & 10.20\% & 17.80\% & 28.60\% & 3.20\%  & 35.20\%  & 15.00\% \\ \hline
\end{tabular}
\caption{Accuracy of Different Models on Various DDoS Attack Types with and without KBs. \label{tab:ddos_results}}
\end{table*}

\section{Simulation and Performance Evaluation}
We use Ollama to retrieve ODLLMs with our constructed KB on Destkop with Nividia RTX 4090 and Intel I9-13900KF \cite{ollama}. We test our model's performance with the CICIoT 2023 dataset \cite{ciciot2023}. Our source code is released on GitHub (https://github.com/claudwq/Intelligent-IoT-Attack-Detection-Design-via-LLM-with-Feature-Ranking-Based-Knowledge-Base-Design.git).
Specifically, we consider the latest small-size models as follows:

 \textbf{Llama 3.1 8B} is the compact variant in the Llama 3.1 series developed by Meta AI with 8 billion parameters \cite{llama3}. 
 %This model is designed for efficient deployment on consumer-grade hardware. Llama 3.1 introduces several enhancements over its predecessors, including an extended context length of up to 128,000 tokens and improved multilingual capabilities, supporting over 30 languages with a primary focus on English. %These advancements enable Llama 3.1 8B to handle more complex tasks and datasets effectively. Its open-source nature allows developers to build and customize AI applications tailored to specific needs. 
\textbf{Phi3 Medium 14B} is part of the Phi-3 series developed by Microsoft with 14 billion parameters, it offers substantial reasoning capabilities while maintaining a manageable computational footprint \cite{phi3}. 
%A notable feature of the Phi-3 models is the use of quantization techniques, which compress the model's weights into lower-precision formats. This approach reduces the overall model size and enhances inference speed and memory efficiency.
%making phi3 Medium 14B suitable for deployment across various devices, including those with limited resources. 
\textbf{Gemma2 9B} is a high-performing and efficient language model within the Gemma 2 series developed by Google DeepMind, which includes models with 2B, 9B, and 27B parameters \cite{gemma2}. 
%The 9B variant is optimized for a balance between performance and resource utilization, making it suitable for various applications. 
%Gemma 2 models are designed to deliver strong performance across diverse tasks while maintaining efficiency, allowing for deployment in environments with varying computational capabilities. 
\textbf{Llama 3.2 3B} is a compact variant in the Llama 3 series with 3 billion parameters, optimized for multilingual tasks and large-scale text processing \cite{llama3}. 
%It strikes a balance between efficiency and accuracy, making it suitable for resource-constrained scenarios while still delivering strong performance across a wide range of applications. 
%Akey feature of Llama 3.2 3B is its support for a context window of up to 128,000 tokens, enabling it to handle extensive text sequences effectively. The model benefits from pretraining on diverse, publicly available datasets and fine-tuning with logits from larger models in the series to enhance its reasoning capabilities.
\textbf{Phi3 Mini 3.8B} is also part of the Phi-3 series with 3.8 billion parameters, designed as a lightweight model optimized for chat-based interactions and reasoning tasks \cite{phi3}. %It provides a high level of performance in tasks requiring conversational understanding and alignment with human preferences. %The model is fine-tuned using supervised fine-tuning (SFT) and Direct Preference Optimization (DPO), ensuring safety and alignment %while maintaining efficiency. Phi3 Mini 3.8B supports context lengths of up to 128,000 tokens, making it effective for processing long text sequences in real-time scenarios.

\subsection{Performance of Medium-size Detectors}

We first evaluate the performance of medium-size models with our KB. As shown in table \ref{tab:ddos_results}, the simulation results evaluate the performance of three ODLLM on detecting four distinct DDoS attack types under three configurations: without a KB (KB), with a long KB, and with a short KB. Accuracy metrics are reported to assess the models’ ability to classify network traffic into these categories.

The long KB is most effective for DDoS-ICMP\_Flood and DDoS-UDP\_Flood detection across all models, particularly for Gemma2 9B, where accuracy increased to 100.00\% for both attacks. However, its performance was inconsistent for other attack types, such as DDoS-TCP\_Flood, where accuracy degraded for Llama 3.1 8B (3.60\%).
The short KB demonstrated a better balance between performance and simplicity, particularly for Phi3 Medium 14B, where accuracy improved for DDoS-UDP\_Flood (59.80\%) and DDoS-PSHACK\_Flood (28.60\%). However, it generally underperformed for Gemma2 9B.

Llama 3.1 8B exhibited the highest accuracy overall, benefiting significantly from both KBs. Its ability to leverage the long KB for DDoS-ICMP\_Flood and the short KB for DDoS-TCP\_Flood highlights its versatility.
Phi3 Medium 14B showed moderate performance, with notable improvement for DDoS-UDP\_Flood and DDoS-PSHACK\_Flood when the short KB was used. This suggests that phi3 Medium 14B is more suited for concise knowledge representation.
Gemma2 9B is heavily reliant on the long KB, achieving perfect accuracy for some attacks but failing entirely for others, such as DDoS-TCP\_Flood.

% DDoS-ICMP\_Flood and DDoS-UDP\_Flood were consistently easier to classify, with high accuracies achieved across all models and configurations.
%  DDoS-TCP\_Flood and DDoS-PSHACK\_Flood posed significant challenges, with low accuracies across most configurations, particularly for Gemma2 9B.

%The simulation results demonstrate that the choice of KB and model significantly impacts the accuracy of DDoS attack detection. 
%Llama 3.1 8B consistently outperformed the other models, with the long KB being particularly effective for certain attack types. 
%However, for resource-constrained environments, the short KB offers a balanced approach, especially when paired with phi3 Medium 14B. 

%Future work will explore optimizing KB representations and improving model fine-tuning to enhance performance for harder-to-detect attacks such as DDoS-TCP\_Flood and DDoS-PSHACK\_Flood.

\begin{table*}[ht]
\centering
\begin{tabular}{|l|c|c|c|c|c|c|}
\hline
\textbf{Attack Type} & \multicolumn{3}{c|}{\textbf{Llama 3.2 3B}} & \multicolumn{3}{c|}{\textbf{Phi3 mini 3.8B}} \\ \hline
                     & \textbf{No KB} & \textbf{With Long KB} & \textbf{With Short KB} & \textbf{No KB} & \textbf{With Long KB} & \textbf{With Short KB} \\ \hline
\textbf{ICMP}   & 28.20\% & 42.00\% & 52.40\% & 6.65\% & 9.60\% & 13.20\% \\ \hline
\textbf{UDP}    & 38.80\% & 23.40\%  & 53.80\% & 0.97\% & 4.20\% & 22.20\% \\ \hline
\textbf{TCP}    & 22.60\% & 27.80\%   & 53.40\% & 0.97\% & 4.20\% & 22.20\%  \\ \hline
\textbf{PSHACK} & 1.60\%  & 3.40\%  & 38.80\% & 0.19\% & 0.00\% & 3.00\% \\ \hline
\end{tabular}
\caption{Accuracy Comparison for DDoS Attack Detection with Different Models and KB Configurations. \label{tab:results}}
\vspace{-0.5cm}
\end{table*}

\subsection{Performance of Small-size Detectors}

To evaluate the effectiveness of the KB configurations on small-size ODLLM, we conducted experiments using two smaller language models: \textbf{Llama 3.2 3B} and \textbf{Phi3 Mini 3.8B}. 
%The models were tested under three KB configurations: without KB, with a long KB, and with a short KB. The datasets included network traffic data representing various DDoS attack types, specifically \textbf{DDoS-ICMP\_Flood}, \textbf{DDoS-UDP\_Flood}, \textbf{DDoS-TCP\_Flood}, and \textbf{DDoS-PSHACK\_Flood}. The goal was to assess the predictive accuracy of each model under different KB configurations.
Table \ref{tab:results} presents the accuracy results for the models across the KB configurations. For Llama 3.2 3B, the model achieved low accuracy without a KB, particularly for DDoS-PSHACK\_Flood, where the accuracy was only 1.60\%. The inclusion of the long KB improved accuracy for DDoS-ICMP\_Flood, achieving 42.00\%, but led to decreased accuracy for DDoS-UDP\_Flood, which dropped to 23.40\%. This suggests that the long KB may have introduced unnecessary complexity, overwhelming the model's capacity. When using the short KB, however, accuracy improved significantly across all attack types. DDoS-UDP\_Flood and DDoS-TCP\_Flood achieved the highest accuracy at 53.80\% and 53.40\%, respectively, demonstrating that short KB retained critical information while reducing complexity.

For Phi3 Mini 3.8B, the model exhibited very low accuracy without a KB, with DDoS-PSHACK\_Flood achieving only 0.19\% accuracy and DDoS-UDP\_Flood reaching just 0.97\%. Adding the long KB resulted in marginal improvements, with DDoS-ICMP\_Flood reaching 9.60\% accuracy and DDoS-UDP\_Flood improving to 4.20\%. However, these results were still significantly lower compared to the short KB configuration. With the short KB, the model's accuracy increased substantially for DDoS-UDP\_Flood and DDoS-TCP\_Flood, both reaching 22.20\%. This demonstrates short KB was far more effective for smaller models.

Both models achieved their highest accuracy for DDoS-ICMP\_Flood across all KB configurations, indicating that the features for this attack type were straightforward and well-represented in the KB. For DDoS-UDP\_Flood, the short KB significantly improved accuracy, particularly for Llama 3.2 3B, which achieved 53.80\%. This highlights the impact of focusing on essential UDP flood features. Similarly, DDoS-TCP\_Flood detection benefitted greatly from the short KB, with Llama 3.2 3B achieving 53.40\% accuracy. Both models struggled to detect DDoS-PSHACK\_Flood, even with the short KB, with Phi3 Mini 3.8B achieving only 3.00\%. This suggests that the feature set for this attack type may require further refinement to enhance detectability.

%The simulation results highlight the critical importance of tailoring KB complexity to the capacity of the model. Smaller models, such as Llama 3.2 3B and Phi3 Mini 3.8B, benefit significantly from a simplified KB that focuses on high-impact features. The short KB configuration consistently outperformed the long KB and no KB configurations, demonstrating that simplifying the feature set enables smaller models to process information effectively and improve predictive accuracy.
%Moreover, ODLLM can also be misled by KB; for example, the performance of Phi3 Medium 14B for DDoS-ICMP\_Flood detection is lower than without KB, which suggests KB design needs to be suitable for the model capability and also can bring potential risks.
%Future work will focus on refining features for complex attack types, such as \textbf{DDoS-PSHACK\_Flood}, to further enhance detection performance.

\section{Conclusions}
In this paper, we presented an intelligent IoT network attack detection framework leveraging ODLLM integrated with feature ranking-based KB designs. The proposed system addresses the challenges of computational resource constraints and data privacy in edge environments while providing a scalable and efficient solution for DDoS attack detection.
Our experiments demonstrated that ODLLMs equipped with a simplified KB tailored to model capacity could achieve competitive performance even on resource-constrained devices. By ranking features using RFR and constructing long and short KBs, we successfully optimized the system's ability to detect various DDoS attack types, including DDoS-ICMP Flood, DDoS-UDP Flood, DDoS-TCP Flood, and DDoS-PSHACK Flood. 

% The results highlight the critical role of tailoring KB complexity to the capacity of the deployed model. While larger models like Llama 3.1 8B benefitted from comprehensive KBs, smaller models such as Llama 3.2 3B and Phi3 Mini 3.8B exhibited significantly better accuracy when using a streamlined KB. This underscores the importance of balancing feature richness and computational efficiency to optimize detection performance.

% Furthermore, our findings indicate that certain attack types, such as DDoS-PSHACK Flood, remain challenging to detect even with optimized KBs. This highlights the need for further refinement of feature selection and KB design to improve accuracy for complex attack patterns. 

% In future work, we plan to explore advanced feature engineering techniques and adaptive KB mechanisms to further enhance detection capabilities. Additionally, extending this framework to include real-time adaptive KB updates and broader attack categories could strengthen IoT network security against emerging threats. 

% Our study provides a foundation for deploying ODLLMs in IoT cybersecurity, offering insights into designing efficient, accurate, and scalable solutions for real-world edge applications.	

\bibliography{references}

\end{document}